\documentclass[reprint,aps,prl,floatfix,superscriptaddress]{revtex4-2}

\usepackage{graphicx}
\usepackage{amsmath}
\usepackage{mathptmx}
\usepackage{ulem}
\usepackage{xcolor}

\begin{document}
\title{Solvated electrons in polar liquids as $\epsilon$-near-zero materials tunable in the terahertz frequency range}
\author{Matthias Runge}
\affiliation{Max-Born-Institut f\"ur Nichtlineare Optik und
	Kurzzeitspektroskopie, 12489 Berlin, Germany}
\email{runge@mbi-berlin.de}
\author{Michael Woerner}
\affiliation{Max-Born-Institut f\"ur Nichtlineare Optik und
	Kurzzeitspektroskopie, 12489 Berlin, Germany}
\author{Denys I. Bondar}
\affiliation{Tulane University, New Orleans, Louisiana 70118, United States}
\author{Thomas Elsaesser}
\affiliation{Max-Born-Institut f\"ur Nichtlineare Optik und
Kurzzeitspektroskopie, 12489 Berlin, Germany}

\begin{abstract}
Electrons in polar liquids give rise to a polaron resonance at a terahertz (THz) frequency $\nu_0$ depending on electron concentration. The impact of this resonance on light propagation is studied in experiments, where a femtosecond pump pulse generates electrons via multiphoton ionization and a THz probe pulse propagated through the excited sample is detected in a phase-resolved way. We observe a behavior characteristic for $\epsilon$-near-zero (ENZ) materials with strongly modified phase and group velocities around $\nu_0$, and a broadening of the THz pulse envelope below $\nu_0$. Calculations based on a local-field approach reproduce the ENZ behavior.
\end{abstract}

\maketitle

$\epsilon$-near-zero (ENZ) materials exhibit unique electromagnetic properties in a frequency range around a zero crossing of the real part of their frequency-dependent electric permittivity $\epsilon(\nu)$, Re$(\epsilon(\nu_0))=0$ \cite{LI17,FO24,GA22}.  Around the frequency $\nu_0$, the peculiar dispersion results in a strong modification of phase and group velocities \cite{JA16A}, which has been exploited for tailoring light propagation on a subwavelength scale \cite{SI06,ED08}, enhancing nonlinear light-matter interaction \cite{CA16,RE19}, and implementing schemes for time refraction of optical pulses \cite{KH20,ZH20,LU23}. Common ENZ materials are bulk and layered solids made from, e.g., Si or transparent conducting oxides, which exhibit ENZ properties close to their plasma frequencies \cite{NA11}. Moreover, a range of optical metamaterials exists, where  ENZ properties are commonly induced by plasmonic nanostructures \cite{MA02H}. The ENZ frequency $\nu_0$ can be tailored by adjusting the doping level of the material or by changing the geometry of nanostructures.  It has been modified by applying an electric bias to a doped sample or by optical pumping as well \cite{PA15,BO21}. 

ENZ behavior in distinct spectral ranges between microwave frequencies and the ultraviolet has mainly been studied by measuring linear optical transmission and/or emission spectra, and by mapping the third-order nonlinear optical response. Time refraction of optical pulses has been demonstrated in propagation experiments, where an optical pump pulse changes the dielectric function during the propagation of a probe pulse and induces a frequency shift of the probe's spectrum \cite{KH20,LU23}. In contrast, propagation experiments directly mapping the impact of ENZ behavior on the phase velocity, group velocity and temporal envelope of optical pulses have remained scarce \cite{GA70,SE85,CH82}. For such studies, the THz frequency range holds particular potential as the electric field of broadband THz pulses can readily be measured in amplitude and phase by electrooptic sampling methods. 

So far, there is a very limited number of ENZ materials for the THz range. A promising system are electrons solvated in polar liquids \cite{TU12B,WE72,KE82,SH95,TA64}. Recent work has demonstrated strong modifications of the THz dielectric function and optical nonlinearity of polar liquids by the introduction of solvated electrons \cite{SA14,GH20A,GH21B,SI22B,RU23A}. In particular, the electrons give rise to a polaronic many-body resonance at a frequency $\nu_0$, where Re$(\epsilon(\nu_0))=0$. The frequency position $\nu_0$ can be tuned by changing the electron concentration. 

In this Letter, we demonstrate the ENZ behavior of a polar liquid containing solvated electrons by mapping the propagation of ultrashort THz pulses in amplitude and phase. 
From such data, we identify an inductive response of the ENZ material causing a pronounced reshaping of the temporal pulse envelopes.
Such results establish solvated electrons in polar liquids as a novel type of ENZ material.

In our current study, the  prototypical alcohol 2-propanol (isopropanol, IPA) serves as the polar solvent. Before presenting experimental results, we analyze the impact of solvated electrons on the dielectric function with the help of a Clausius-Mossotti  approach \cite{HA83}. This model treats electrons and solvent molecules as point-like dipoles, interacting with each other via the local electric field. The dielectric function is given by the expression \cite{GH21B,WO22}
\begin{eqnarray}
	3\dfrac{\epsilon(\nu,c_e)-1}{\epsilon(\nu,c_e)+2}&=&3\dfrac{\epsilon_\text{neat}(\nu)-1}{\epsilon_\text{neat}(\nu)+2} + c_eN_A\alpha_\text{el}(\nu),
	\label{eq:ClausiusMossotti}\\
	\text{with  }\alpha_\text{el}&=&-\dfrac{e^2}{\epsilon_0m[(2\pi\nu)^2+i\gamma(2\pi\nu)]},
	\label{eq:ElectronPolarization}
\end{eqnarray}
with Avogadro's constant $N_A$, the polarizability of free electrons $\alpha_{el}$,  the elementary charge $e$, the vacuum permittivity $\epsilon_0$, the electronic mass $m$, and a damping constant $\gamma$, set to zero in the calculations. The quantity $\epsilon(\nu, c_e) = \epsilon^\prime (\nu,c_e) + i \epsilon^{\prime \prime} (\nu, c_e)$ is the complex dielectric function of the liquid containing solvated electrons of a concentration $c_e$, while $\epsilon_\text{neat} (\nu)$ represents the dielectric function of the neat solvent.

\begin{figure}[t!]
	\begin{center}
		\includegraphics[width=0.5\textwidth]{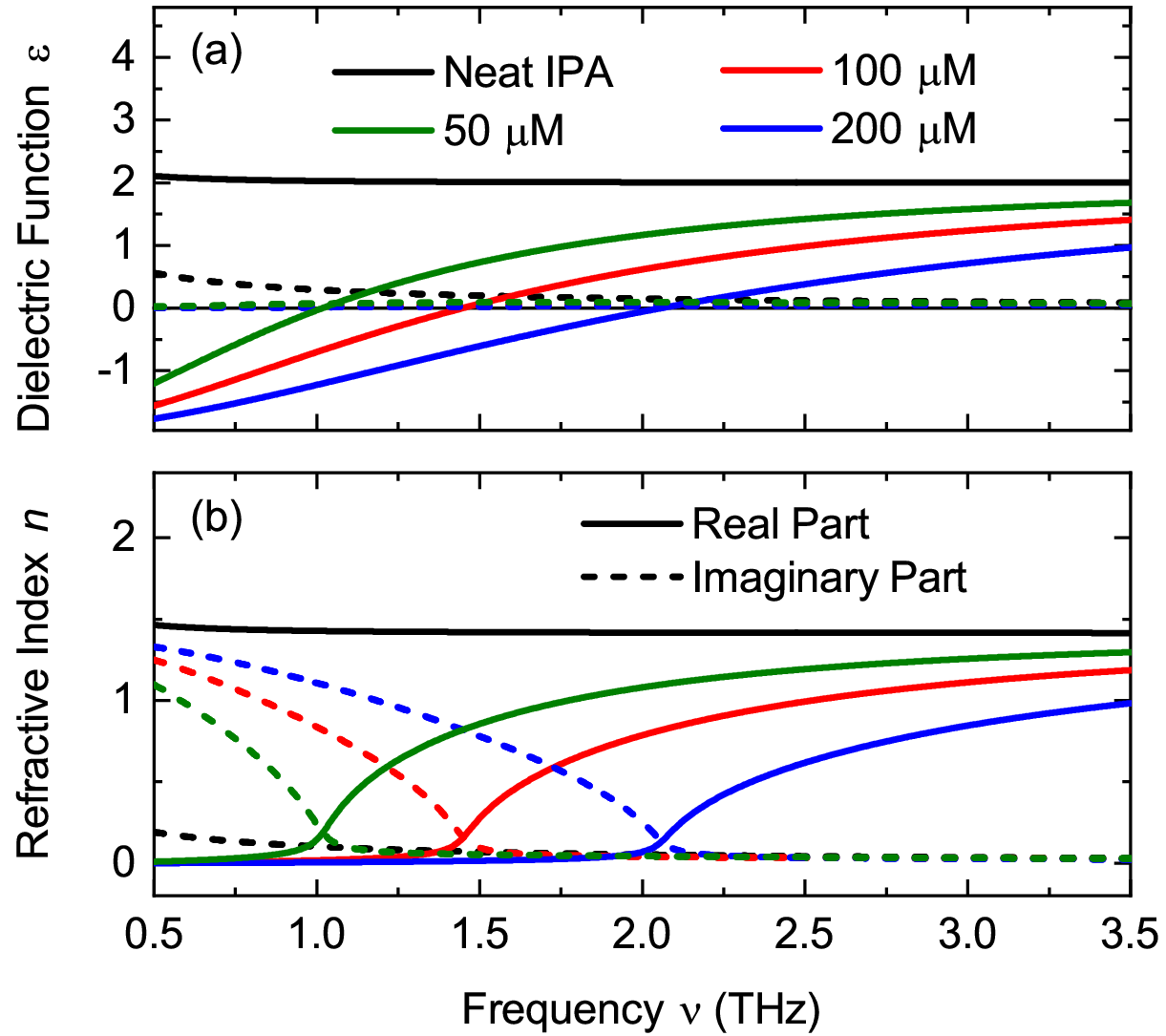}
	\end{center}
	\caption{(a)~Complex dielectric function of neat IPA (black curves) as extracted from numerical fits considering a single Debye band to linear time-domain THz spectra \cite{RU23A} and of IPA including certain concentrations of solvated electrons of $c_e=50$, 100 and 200~$\mu$M (green, red and blue curves) as calculated from the Clausius-Mossotti equation. (b)~Real and imaginary parts of the refractive index $n=\sqrt{\epsilon}$ of neat IPA and IPA including solvated electrons, calculated from the complex dielectric functions in panel~(a). }   
	\label{fig:TH_eps}
\end{figure}

Figure~\ref{fig:TH_eps}(a) shows the dielectric function $\epsilon_\text{neat}(\nu)$ of neat IPA (black lines) as a function of frequency $\nu$ compared to the dielectric function $\epsilon(\nu,c_e)$ of IPA including different concentrations $c_e$ of solvated electrons (colored lines). The presence of electrons modifies the flat frequency dispersion of Re$(\epsilon_\text{neat} (\nu)) = \epsilon^\prime_\text{neat} (\nu)$ strongly and, in particular, generates a zero crossing at a frequency position $\nu_0$ depending on the electron concentration $c_e$. Im$(\epsilon_\text{neat} (\nu))$ is modified as well but maintains its comparably small values, resulting in weak absorption losses. Correspondingly, the real part of the refractive index $n = \sqrt{\epsilon}$ approaches the zero line at $\nu_0$ and remains close to zero at lower frequencies [Fig. 1(b)]. It is important to note that $\text{Re}(n) \gg \text{Im}(n)$ for $\nu > \nu_0$, while $\text{Im}(n) \gg \text{Re}(n)$ for $\nu < \nu_0$. This behavior suggests that the material response for $\nu>\nu_0$ is dominated by dispersion  and for $\nu<\nu_0$ by absorption, leading to vastly different phase and group velocities in the two regimes - a main characteristic of ENZ materials.

\begin{figure*}[t!]
	\begin{center}
		\includegraphics[width=1\textwidth]{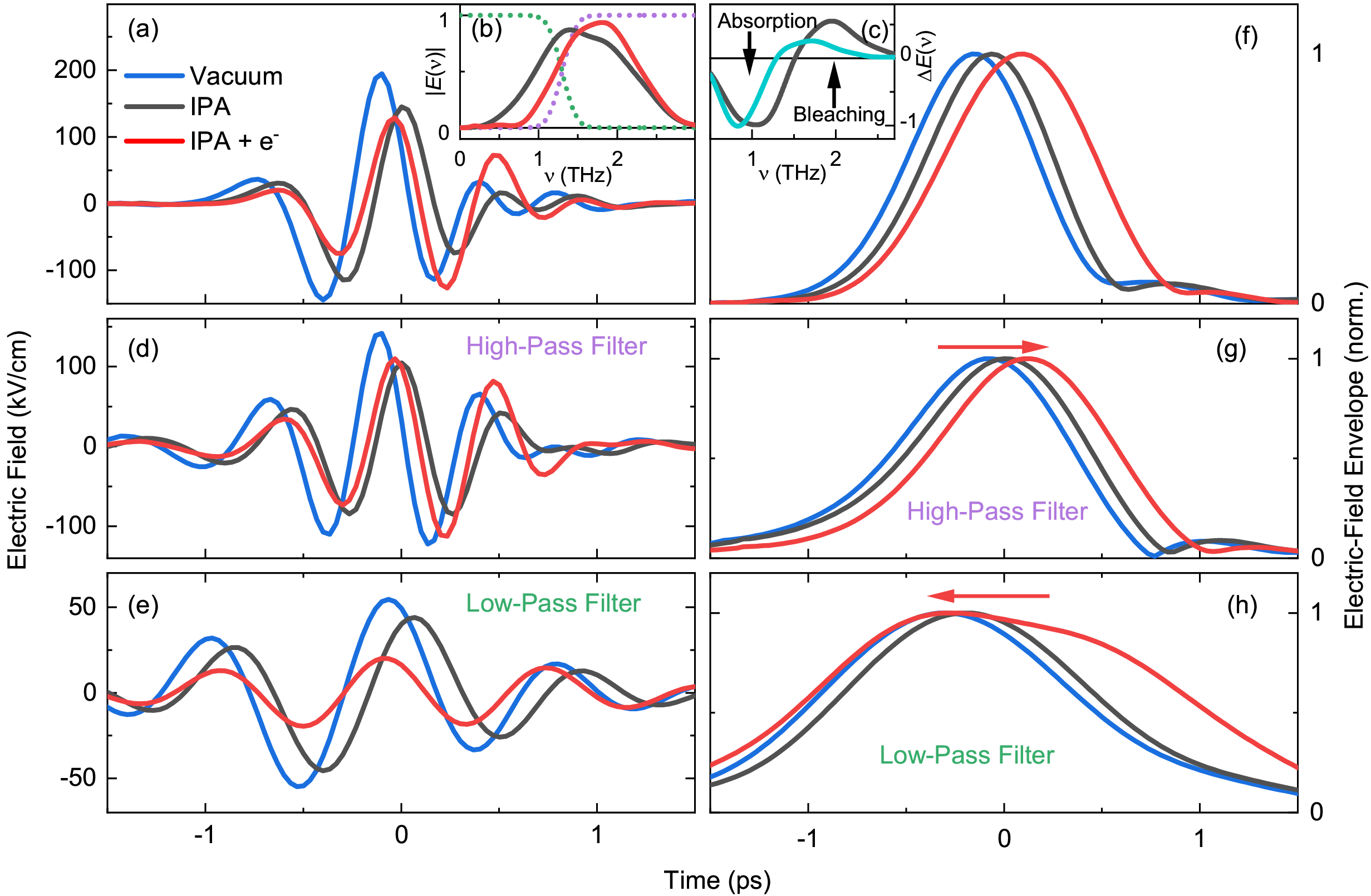}
	\end{center}
	\caption{(a)~THz transients measured in vacuum (blue curve), propagating through neat IPA (black curve), and through IPA including a concentration of solvated electrons of $c_e=100~\mu$M (red curve). The latter transient was corrected by a frequency dependent transfer function estimated from the transmission through a 100~$\mu$m wide pinhole. 
	(b)~Spectrum of the THz pulses transmitted through neat IPA and through IPA including solvated electrons obtained from Fourier transforming the transients in panel~(a). The green and the purple dotted lines indicate filter functions for low- and high pass filtering of the data. (c)~Difference spectra $|E_\text{Pr}^\text{Pu}(\nu)|-|E_\text{Pr}(\nu)|$ for electron concentrations of 100 and 75~$\mu$M (black and cyan lines). (d,~e)~THz transients filtered with a high- and a low-pass Fourier filter [see panel~(b)]. (f,~g,~h)~Electric-field envelopes of the THz pulses in panels~(a),~(d) and~(e).  }   
	\label{fig:Exp_uebersicht}
\end{figure*}

The experiments were performed on a liquid IPA jet with a thickness of $s=75~\mu$m. Solvated electrons  are generated by five- to six-photon absorption using a pump pulse of a center wavelength of 800 nm, a duration of 30 ~fs, and a pulse energy of  300~$\mu$J, generated in a Ti:sapphire oscillator-amplifier system \cite{GH20A,GH21B,SI22B,RU23A}. 
The pump pulse is focused onto the sample jet with a  lens of a 300-mm focal length, leading to a spot size of 250~$\mu$m on the sample. Due to the multi-photon character of electron generation, the area in which electrons are generated, is smaller with a diameter of some 100~$\mu$m.
The electron concentration of about 100~$\mu$M is derived from a measurement of the absorption of an 800-nm probe pulse in the excited sample and the molar extinction coefficient of solvated electrons $\epsilon = 1.388 \times 10^4$~M$^{-1}$cm$^{-1}$ at 800~nm \cite{SA65}. Details are given in the Supplemental Material (SM).

Changes of the THz dielectric function of IPA including solvated electrons are probed at a time delay of 3~ps after electron generation with a single few-cycle THz pulse with a peak electric field of about 100 kV/cm and a spectrum between 0.5 and 3~THz.  The THz pulse is focused onto the sample using an off-axis parabolic mirror (focal length 25.4~mm) giving a diffraction-limited spot size of about 300~$\mu$m. The THz pulse transmitted through the sample is detected by electrooptic sampling, providing the time dependent amplitude and phase of the THz electric field. A ZnTe crystal (thickness 10~$\mu$m) serves as the electrooptic medium and 10-fs pulses from the oscillator of the Ti:sapphire laser system as readout pulses \cite{RE21}. Since the pumped area on the sample is smaller than the cross section of the THz probe beam, the data is corrected for this mismatch by a frequency dependent transfer function obtained from the THz transmission through a 100~$\mu$m-wide pinhole. Details of this procedure are given in the SM.

Figure~\ref{fig:Exp_uebersicht} summarizes the results of the time-resolved THz experiments. In Fig.~\ref{fig:Exp_uebersicht}(a), the   electric field of the THz pulses traveling in vacuum (sample jet switched off, blue line), transmitted through neat IPA (black line), and transmitted through IPA containing electrons (red line) is plotted as a function of time. The electron concentration of 
$c_e \approx 100~\mu$M corresponds to a zero crossing of Re$(\epsilon)$ at $\nu_0 = 1.4$~THz (cf. Fig.~\ref{fig:TH_eps}.
The maximum of the THz field $E_\text{Pr}(t)$ transmitted through the neat IPA sample is delayed by about 105~fs compared to the pulse traveling in vacuum. This delay is consistent with the retardation $\Delta t = \Delta n s/c_0 = 110$~fs, estimated from the jet thickness $s$ and the difference $\Delta n$ in refractive index between IPA and vacuum ($c_0$: vacuum speed of light). In contrast, the maximum of the THz field $E_\text{Pr}^\text{Pu}(t)$ transmitted through IPA including solvated electrons appears somewhat earlier than without electrons. Furthermore, the electric field at late times is enhanced, reflecting a temporal broadening of the pulse. 

A Fourier transform of the THz transients $E_\text{Pr}(t)$ and $E_\text{Pr}^\text{Pu}(t)$  gives the spectra $|E_\text{Pr}(\nu)|$ and $|E_\text{Pr}^\text{Pu}(\nu)|$ for neat IPA and IPA including solvated electrons, as presented in Fig.~\ref{fig:Exp_uebersicht}(b). For frequencies $\nu<\nu_0 = 1.5$~THz, $|E^\text{Pu}_\text{Pr}(\nu)|$ is smaller than $|E_\text{Pr}(\nu)|$, and for frequencies $\nu> \nu_0$, $|E^\text{Pu}_\text{Pr}(\nu)|$ larger than $|E_\text{Pr}(\nu)|$. The latter aspect becomes most clear when calculating the difference spectrum $\Delta E(\nu)=|E_\text{Pr}^\text{Pu}(\nu)|-|E_\text{Pr}(\nu)|$ [black line in Fig.~\ref{fig:Exp_uebersicht}(c)], which shows a transition from negative values (absorption) to positive values (bleaching). The frequency at which this transition occurs is tunable by changing the electron concentration as demonstrated by a measurement with a lower electron concentration of $75~\mu$M, leading to a transition from absorption to bleaching at $\nu_0=1.2~$THz (cyan line).


To compare the frequency regions below and above $\nu_0$, we Fourier filtered the time domain data selecting low- and high-frequency components by convoluting the spectra with a step function [green and purple lines in Fig.~\ref{fig:Exp_uebersicht}(b)], and then performing an inverse Fourier transform to the time domain. The results of this procedure are presented in Figs.~\ref{fig:Exp_uebersicht}(d,~e). 
For the high-frequency transients, the THz pulses transmitted through the sample including electrons (solid red line) appears slightly earlier than in the case without electrons (black line). The amplitude of the THz transients is mainly unchanged, as also expected from the very small imaginary part of the refractive index at high frequencies  [cf. Fig.~\ref{fig:TH_eps}(b)]. 

In the frequency range below $\nu_0$, the THz transient transmitted through IPA including electrons appears substantially earlier and has a distinctively decreased amplitude compared to the transient transmitted through neat IPA.  The THz electric field transmitted through IPA including electrons has a phase similar to the transient propagated in vacuum, a behavior pointing to an average phase velocity close to the vacuum speed of light. As shown in Fig. 1(b), the imaginary part of the refractive index Im($n$) dominates at $\nu < \nu_0$, leading to appreciable absorption losses. Correspondingly, the amplitude of the transmitted THz transient is reduced by about 40\% compared to the transient propagating in vacuum.

THz pulse envelopes, which propagate with the group velocity $v_g$, were derived from the phase-resolved electric field transients (for details, see SM). The envelopes of the unfiltered THz transients are plotted in Fig.~\ref{fig:Exp_uebersicht}(f). The  envelope transmitted through IPA including electrons is delayed by about 100~fs compared to the envelope of the THz pulse transmitted through neat IPA, indicating a decrease in group velocity due to the presence of electrons. The shapes of the different envelopes and their temporal widths are similar. Figures~\ref{fig:Exp_uebersicht}(g,~h) display the normalized envelopes of the high- and low-frequency THz transients. The high-frequency envelope [Fig.~\ref{fig:Exp_uebersicht}(g)] shows a behavior similar to the unfiltered envelopes, i.e., the overall THz response is dominated by the high-frequency contribution. In contrast, the low-frequency envelope transmitted through IPA including electrons [Fig. ~\ref{fig:Exp_uebersicht}(h)] shows a shift toward earlier times and a pronounced temporal broadening. While the leading edge of the THz pulse (at negative times) coincides with the THz pulse propagated in vacuum, the electric field re-emitted by the THz polarization of the sample leads to the pronounced temporal broadening of the envelope (see discussion below).

The temporal shifts of the time-resolved electric-field transients [Fig.~\ref{fig:Exp_uebersicht}(a,~d,~e)] reflect changes in phase velocity $v_p$ upon introducing solvated electrons, while the pulse envelopes propagate with the group velocity $v_g$. Relevant are the phase and group  velocities averaged over the spectrum of the broadband THz pulse [cf. Fig.~\ref{fig:Exp_uebersicht}(b)]. Below $\nu_0$, where Re$(n) \approx 0$, the pronounced imaginary part Im$(n)$ of the refractive index leads to a reshaping of the THz transient. Absorption generates a time-dependent current in the sample, which acts as a source of THz radiation and contributes to the total THz field transmitted through the sample. This `inductive' response is mainly relevant for the trailing part of the THz pulses, causing a reshaping of the pulse envelope for which the standard definition of the group velocity $v_g=d\omega/dk$ is inappropriate.

\begin{figure}[t!]
	\begin{center}
		\includegraphics[width=0.48\textwidth]{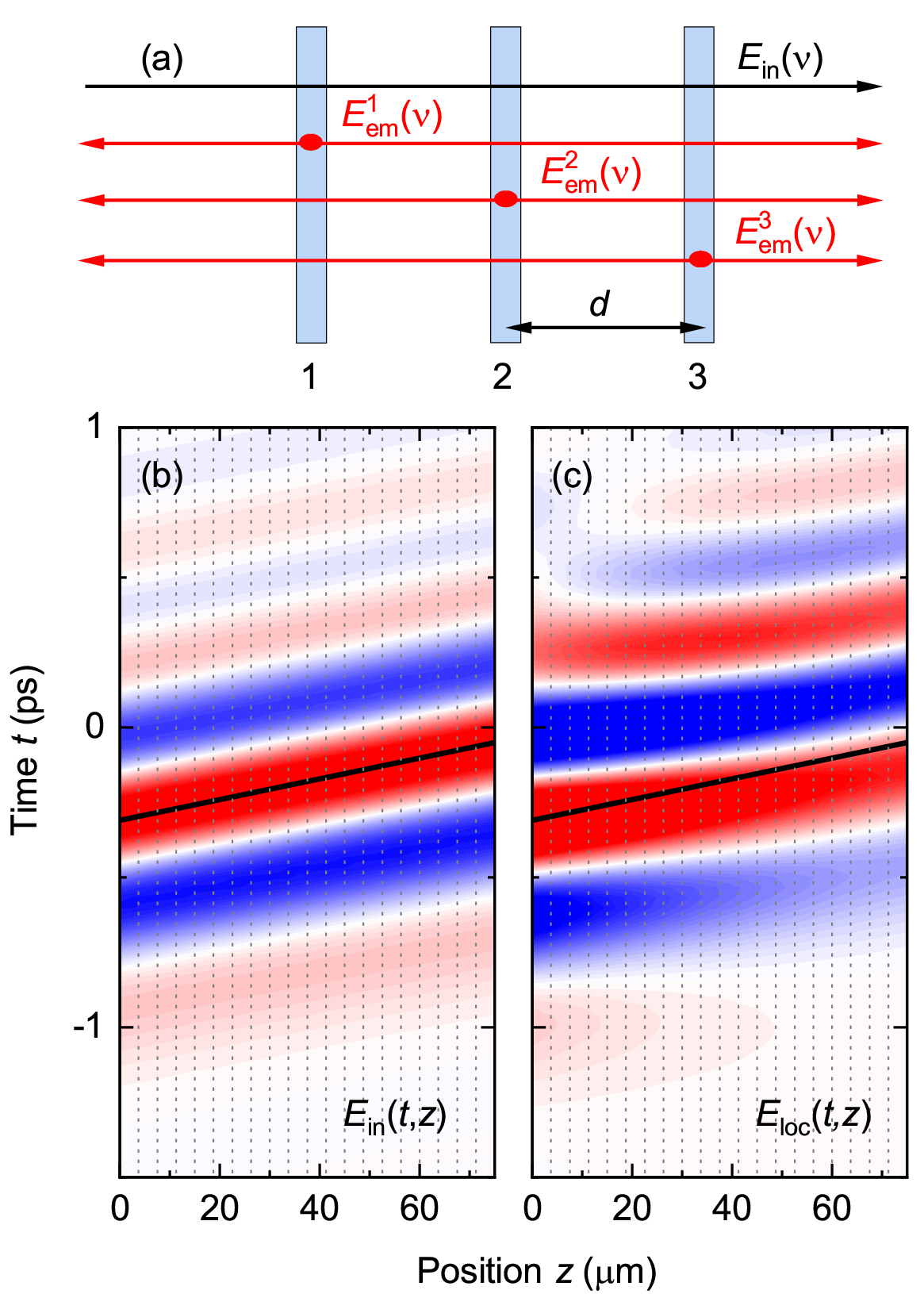}
	\end{center}
	\caption{(a)~Electric fields in a multi-layer geometry with three layers, in which the local electric field $E_\text{loc}(\nu)$ in each layer consists of the interference of the incident electric field $E_\text{in}(\nu)$ and all the emitted electric fields $E_\text{em}^k(\nu)$. (b,~c)~Contour plot of the incident electric field $E_\text{in}(t)$ and the local electric field $E_\text{loc}(t)$ as a function of time $t$ and position $z$ inside the sample.}   
	\label{fig:multilayer}
\end{figure}

For a theoretical description of propagation of the THz pulse shown in Fig.~\ref{fig:Exp_uebersicht}(a) (blue line), the emitted electric field $E_\text{em}(\nu)$ as a function of frequency is calculated from the current $j(\nu)$ induced by the local electric field $E_\text{loc}(\nu)$ in the sample, using 
\begin{eqnarray}
	E_\text{em}(\nu)=&& \dfrac{s}{2\epsilon_0 c}j(\nu)=-\dfrac{s}{2\epsilon_0 c}\sigma(\nu)E_\text{loc}(\nu) \label{eq:Eem}\\
	&&\text{with } 	\sigma(\nu)=-2\pi i\nu\epsilon_0\cdot[\epsilon(\nu)-1],
\end{eqnarray}
where $\sigma(\nu)$ is the frequency-dependent conductivity and $s$ the thickness. Eq.~\eqref{eq:Eem} is an appropriate approximation  for a sample thin compared to the wavelength, i.e., $s\ll\lambda$. In a scenario where $s\approx\lambda$, the sample is conceptually decomposed into $n$ thin layers with a thickness of $d=s/n\ll\lambda$, where the electric field emitted from each individual layer is radiatively coupled to all the other layers in the sample, as schematically shown for the case of $n=3$ in Fig.~\ref{fig:multilayer}(a) \cite{SH05A}. Thus, in each of these layers, the undistorted  incident electric field $E_\text{in}(\nu)$ and the electric fields $E^l_\text{em}(\nu)$ emitted from the other layers are superimposed to a local electric field given by
\begin{equation}
	E_\text{loc}^j(\nu)=E_\text{in}(\nu)e^{ikdj}+ \sum_{l=0}^nE_\text{em}^l(\nu)e^{ikd|l-j|}, \label{eq:Eloc}
\end{equation}
with $k=\frac{2\pi\nu}{c}$. The electric field transmitted through the sample corresponds to the local electric field in the $n$-th layer, i.e., $E_\text{tr}(\nu)=E_\text{loc}^n(\nu)$, whereas the reflected electric field is given by $E_\text{re}(\nu)=E_\text{loc}^1(\nu)-E_\text{in}(\nu)$. The electric fields $E_\text{loc}^j(t)$ in the time domain are obtained by performing inverse Fourier transforms along $\nu$.

Figures~\ref{fig:multilayer}(b,~c) display the incident electric field $E_\text{in}(t,z)$ and the local electric field $E_\text{loc}(t,z)$ calculated from Eq.~\eqref{eq:Eloc} as functions of time $t$ and position $z$. $E_\text{in}(t,z)$ propagates through vacuum whereas $E_\text{loc}(t,z)$ propagates through  liquid IPA including a concentration of electrons of $c_e=100~\mu$M, as considered by using the dielectric function calculated from the Clausius-Mossotti equation in Eq.~\eqref{eq:ClausiusMossotti}. In the calculations, the sample with a thickness of $s=75~\mu$m is separated into $n=20$ individual thin films  (grey dotted lines). The temporal shape of $E_\text{in}(t,z)$ does not change during propagation in vacuum. The thick black line indicates the maximum position of the electric field, with a slope corresponding to the inverse vacuum speed of light $1/c$.  In contrast, $E_\text{loc}(t,z)$ shows distinct modifications in the different layers inside the sample. The maximum of $E_\text{loc}(t,z)$ has a smaller slope than $E_\text{in}(t,z)$ as indicated by the black line, i.e., the phase velocity exceeds the vacuum speed of light, which is a clear hallmark of the ENZ character already observed in the experiments. Simultaneously, an energy redistribution across the THz pulse occurs. In the first layers of the sample, $E_\text{loc}(t,z)$ shows pronounced contributions at early times around $t=-1$~ps which never exceed the electric field of the incident pulse since energy cannot be transported faster than the vacuum speed of light. These contributions decrease with increasing $z$, while new contributions at later times $t>0.5$~ps arise, leading to a tail and temporal broadening as also observed in the experimental data in Fig.~\ref{fig:Exp_uebersicht}. This tail arises from the radiative coupling of electric currents in the different layers inside the sample.

We conclude that electrons solvated in polar liquids represent a novel type of ENZ materials. The ENZ behavior stems from strong modifications to the THz dielectric function, which exhibits a zero transition after adding solvated electrons. A THz pulse propagating through this medium has vastly different phase and group velocities at frequencies below and above this zero transition, an effect characteristic to ENZ materials. While the present study focuses on linear pulse propagation, the optical nonlinearity introduced by solvated electrons may allow for implementing fully functional ENZ materials that are optically controllable in a wide range of THz frequencies. Our results are related to a recent theoretical finding that appropriately temporally-shaped laser pulses can drive an ENZ-like response from any system \cite{MA23}.

\begin{acknowledgments}
This research has received funding from the European Research Council (ERC) under the European Union’s Horizon 2020 Research and Innovation Program (grant agreement 833365). D.I.B. was supported by Army Research Office (ARO) (grant W911NF-23-1-0288; program manager Dr.~James Joseph). The views and conclusions contained in this document are those of the authors and should not be interpreted as representing the official policies, either expressed or implied, of ARO or the U.S. Government. The U.S. Government is authorized to reproduce and distribute reprints for Government purposes notwithstanding any copyright notation herein.
\end{acknowledgments}


\begin{thebibliography}{31}%
	\makeatletter
	\providecommand \@ifxundefined [1]{%
		\@ifx{#1\undefined}
	}%
	\providecommand \@ifnum [1]{%
		\ifnum #1\expandafter \@firstoftwo
		\else \expandafter \@secondoftwo
		\fi
	}%
	\providecommand \@ifx [1]{%
		\ifx #1\expandafter \@firstoftwo
		\else \expandafter \@secondoftwo
		\fi
	}%
	\providecommand \natexlab [1]{#1}%
	\providecommand \enquote  [1]{``#1''}%
	\providecommand \bibnamefont  [1]{#1}%
	\providecommand \bibfnamefont [1]{#1}%
	\providecommand \citenamefont [1]{#1}%
	\providecommand \href@noop [0]{\@secondoftwo}%
	\providecommand \href [0]{\begingroup \@sanitize@url \@href}%
	\providecommand \@href[1]{\@@startlink{#1}\@@href}%
	\providecommand \@@href[1]{\endgroup#1\@@endlink}%
	\providecommand \@sanitize@url [0]{\catcode `\\12\catcode `\$12\catcode
		`\&12\catcode `\#12\catcode `\^12\catcode `\_12\catcode `\%12\relax}%
	\providecommand \@@startlink[1]{}%
	\providecommand \@@endlink[0]{}%
	\providecommand \url  [0]{\begingroup\@sanitize@url \@url }%
	\providecommand \@url [1]{\endgroup\@href {#1}{\urlprefix }}%
	\providecommand \urlprefix  [0]{URL }%
	\providecommand \Eprint [0]{\href }%
	\providecommand \doibase [0]{https://doi.org/}%
	\providecommand \selectlanguage [0]{\@gobble}%
	\providecommand \bibinfo  [0]{\@secondoftwo}%
	\providecommand \bibfield  [0]{\@secondoftwo}%
	\providecommand \translation [1]{[#1]}%
	\providecommand \BibitemOpen [0]{}%
	\providecommand \bibitemStop [0]{}%
	\providecommand \bibitemNoStop [0]{.\EOS\space}%
	\providecommand \EOS [0]{\spacefactor3000\relax}%
	\providecommand \BibitemShut  [1]{\csname bibitem#1\endcsname}%
	\let\auto@bib@innerbib\@empty
	\bibitem [{\citenamefont {Liberal}\ and\ \citenamefont {Engheta}(2017)}]{LI17}%
	\BibitemOpen
	\bibfield  {author} {\bibinfo {author} {\bibfnamefont {I.}~\bibnamefont
			{Liberal}}\ and\ \bibinfo {author} {\bibfnamefont {N.}~\bibnamefont
			{Engheta}},\ }\bibfield  {title} {\bibinfo {title} {Near-zero refractive
			index photonics},\ }\href@noop {} {\bibfield  {journal} {\bibinfo  {journal}
			{Nat. Photon.}\ }\textbf {\bibinfo {volume} {11}},\ \bibinfo {pages} {149}
		(\bibinfo {year} {2017})}\BibitemShut {NoStop}%
	\bibitem [{\citenamefont {Fomra}\ \emph {et~al.}(2024)\citenamefont {Fomra},
		\citenamefont {Ball}, \citenamefont {Saha}, \citenamefont {Wu}, \citenamefont
		{Sojib}, \citenamefont {Agrawal}, \citenamefont {Lezec},\ and\ \citenamefont
		{Kinsey}}]{FO24}%
	\BibitemOpen
	\bibfield  {author} {\bibinfo {author} {\bibfnamefont {D.}~\bibnamefont
			{Fomra}}, \bibinfo {author} {\bibfnamefont {A.}~\bibnamefont {Ball}},
		\bibinfo {author} {\bibfnamefont {S.}~\bibnamefont {Saha}}, \bibinfo {author}
		{\bibfnamefont {J.}~\bibnamefont {Wu}}, \bibinfo {author} {\bibfnamefont
			{M.}~\bibnamefont {Sojib}}, \bibinfo {author} {\bibfnamefont
			{A.}~\bibnamefont {Agrawal}}, \bibinfo {author} {\bibfnamefont {H.~J.}\
			\bibnamefont {Lezec}},\ and\ \bibinfo {author} {\bibfnamefont
			{N.}~\bibnamefont {Kinsey}},\ }\bibfield  {title} {\bibinfo {title}
		{Nonlinear optics at epsilon near zero: From origins to new materials},\
	}\href@noop {} {\bibfield  {journal} {\bibinfo  {journal} {Appl. Phys. Rev.}\
		}\textbf {\bibinfo {volume} {11}},\ \bibinfo {pages} {011317} (\bibinfo
		{year} {2024})}\BibitemShut {NoStop}%
	\bibitem [{\citenamefont {Galiffi}\ \emph {et~al.}(2022)\citenamefont
		{Galiffi}, \citenamefont {Tirole}, \citenamefont {Yin}, \citenamefont {Li},
		\citenamefont {Vezzoli}, \citenamefont {Huidobro}, \citenamefont
		{Silveirinha}, \citenamefont {Sapienza}, \citenamefont {Alu},\ and\
		\citenamefont {Pendry}}]{GA22}%
	\BibitemOpen
	\bibfield  {author} {\bibinfo {author} {\bibfnamefont {E.}~\bibnamefont
			{Galiffi}}, \bibinfo {author} {\bibfnamefont {R.}~\bibnamefont {Tirole}},
		\bibinfo {author} {\bibfnamefont {S.}~\bibnamefont {Yin}}, \bibinfo {author}
		{\bibfnamefont {H.}~\bibnamefont {Li}}, \bibinfo {author} {\bibfnamefont
			{S.}~\bibnamefont {Vezzoli}}, \bibinfo {author} {\bibfnamefont {P.~A.}\
			\bibnamefont {Huidobro}}, \bibinfo {author} {\bibfnamefont {M.~G.}\
			\bibnamefont {Silveirinha}}, \bibinfo {author} {\bibfnamefont
			{R.}~\bibnamefont {Sapienza}}, \bibinfo {author} {\bibfnamefont
			{A.}~\bibnamefont {Alu}},\ and\ \bibinfo {author} {\bibfnamefont {J.~B.}\
			\bibnamefont {Pendry}},\ }\bibfield  {title} {\bibinfo {title} {Photonics of
			time-varying media},\ }\href@noop {} {\bibfield  {journal} {\bibinfo
			{journal} {Adv. Photonics}\ }\textbf {\bibinfo {volume} {4}},\ \bibinfo
		{pages} {014002} (\bibinfo {year} {2022})}\BibitemShut {NoStop}%
	\bibitem [{\citenamefont {Javani}\ and\ \citenamefont
		{Stockman}(2016)}]{JA16A}%
	\BibitemOpen
	\bibfield  {author} {\bibinfo {author} {\bibfnamefont {M.~H.}\ \bibnamefont
			{Javani}}\ and\ \bibinfo {author} {\bibfnamefont {M.~I.}\ \bibnamefont
			{Stockman}},\ }\bibfield  {title} {\bibinfo {title} {Real and imaginary
			properties of epsilon-near-zero materials},\ }\href
	{https://doi.org/10.1103/physrevlett.117.107404} {\bibfield  {journal}
		{\bibinfo  {journal} {Phys. Rev. Lett.}\ }\textbf {\bibinfo {volume} {117}},\
		\bibinfo {pages} {107404} (\bibinfo {year} {2016})}\BibitemShut {NoStop}%
	\bibitem [{\citenamefont {Silveirinha}\ and\ \citenamefont
		{Engheta}(2006)}]{SI06}%
	\BibitemOpen
	\bibfield  {author} {\bibinfo {author} {\bibfnamefont {M.}~\bibnamefont
			{Silveirinha}}\ and\ \bibinfo {author} {\bibfnamefont {N.}~\bibnamefont
			{Engheta}},\ }\bibfield  {title} {\bibinfo {title} {{Tunneling of
				Electromagnetic Energy through Subwavelength Channels and Bends using}
			$\epsilon$-{Near-Zero Materials}},\ }\href
	{https://doi.org/10.1103/physrevlett.97.157403} {\bibfield  {journal}
		{\bibinfo  {journal} {Phys. Rev. Lett.}\ }\textbf {\bibinfo {volume} {97}},\
		\bibinfo {pages} {157403} (\bibinfo {year} {2006})}\BibitemShut {NoStop}%
	\bibitem [{\citenamefont {Edwards}\ \emph {et~al.}(2008)\citenamefont
		{Edwards}, \citenamefont {Alù}, \citenamefont {Young}, \citenamefont
		{Silveirinha},\ and\ \citenamefont {Engheta}}]{ED08}%
	\BibitemOpen
	\bibfield  {author} {\bibinfo {author} {\bibfnamefont {B.}~\bibnamefont
			{Edwards}}, \bibinfo {author} {\bibfnamefont {A.}~\bibnamefont {Alù}},
		\bibinfo {author} {\bibfnamefont {M.~E.}\ \bibnamefont {Young}}, \bibinfo
		{author} {\bibfnamefont {M.}~\bibnamefont {Silveirinha}},\ and\ \bibinfo
		{author} {\bibfnamefont {N.}~\bibnamefont {Engheta}},\ }\bibfield  {title}
	{\bibinfo {title} {Experimental verification of epsilon-near-zero
			metamaterial coupling and energy squeezing using a microwave waveguide},\
	}\href {https://doi.org/10.1103/physrevlett.100.033903} {\bibfield  {journal}
		{\bibinfo  {journal} {Phys. Rev. Lett.}\ }\textbf {\bibinfo {volume} {100}},\
		\bibinfo {pages} {033903} (\bibinfo {year} {2008})}\BibitemShut {NoStop}%
	\bibitem [{\citenamefont {Caspani}\ \emph {et~al.}(2016)\citenamefont
		{Caspani}, \citenamefont {Kaipurath}, \citenamefont {Clerici}, \citenamefont
		{Ferrera}, \citenamefont {Roger}, \citenamefont {Kim}, \citenamefont
		{Kinsey}, \citenamefont {Pietrzyk}, \citenamefont {Di~Falco}, \citenamefont
		{Shalaev}, \citenamefont {Boltasseva},\ and\ \citenamefont {Faccio}}]{CA16}%
	\BibitemOpen
	\bibfield  {author} {\bibinfo {author} {\bibfnamefont {L.}~\bibnamefont
			{Caspani}}, \bibinfo {author} {\bibfnamefont {R.}~\bibnamefont {Kaipurath}},
		\bibinfo {author} {\bibfnamefont {M.}~\bibnamefont {Clerici}}, \bibinfo
		{author} {\bibfnamefont {M.}~\bibnamefont {Ferrera}}, \bibinfo {author}
		{\bibfnamefont {T.}~\bibnamefont {Roger}}, \bibinfo {author} {\bibfnamefont
			{J.}~\bibnamefont {Kim}}, \bibinfo {author} {\bibfnamefont {N.}~\bibnamefont
			{Kinsey}}, \bibinfo {author} {\bibfnamefont {M.}~\bibnamefont {Pietrzyk}},
		\bibinfo {author} {\bibfnamefont {A.}~\bibnamefont {Di~Falco}}, \bibinfo
		{author} {\bibfnamefont {V.}~\bibnamefont {Shalaev}}, \bibinfo {author}
		{\bibfnamefont {A.}~\bibnamefont {Boltasseva}},\ and\ \bibinfo {author}
		{\bibfnamefont {D.}~\bibnamefont {Faccio}},\ }\bibfield  {title} {\bibinfo
		{title} {Enhanced nonlinear refractive index in $\epsilon$-near-zero
			materials},\ }\href {https://doi.org/10.1103/physrevlett.116.233901}
	{\bibfield  {journal} {\bibinfo  {journal} {Phys. Rev. Lett.}\ }\textbf
		{\bibinfo {volume} {116}},\ \bibinfo {pages} {233901} (\bibinfo {year}
		{2016})}\BibitemShut {NoStop}%
	\bibitem [{\citenamefont {Reshef}\ \emph {et~al.}(2019)\citenamefont {Reshef},
		\citenamefont {De~Leon}, \citenamefont {Alam},\ and\ \citenamefont
		{Boyd}}]{RE19}%
	\BibitemOpen
	\bibfield  {author} {\bibinfo {author} {\bibfnamefont {O.}~\bibnamefont
			{Reshef}}, \bibinfo {author} {\bibfnamefont {I.}~\bibnamefont {De~Leon}},
		\bibinfo {author} {\bibfnamefont {M.~Z.}\ \bibnamefont {Alam}},\ and\
		\bibinfo {author} {\bibfnamefont {R.~W.}\ \bibnamefont {Boyd}},\ }\bibfield
	{title} {\bibinfo {title} {Nonlinear optical effects in epsilon-near-zero
			media},\ }\href {https://doi.org/10.1038/s41578-019-0120-5} {\bibfield
		{journal} {\bibinfo  {journal} {Nat. Rev. Mat.}\ }\textbf {\bibinfo {volume}
			{4}},\ \bibinfo {pages} {535} (\bibinfo {year} {2019})}\BibitemShut {NoStop}%
	\bibitem [{\citenamefont {Khurgin}\ \emph {et~al.}(2020)\citenamefont
		{Khurgin}, \citenamefont {Clerici}, \citenamefont {Bruno}, \citenamefont
		{Caspani}, \citenamefont {DeVault}, \citenamefont {Kim}, \citenamefont
		{Shaltout}, \citenamefont {Boltasseva}, \citenamefont {Shalaev},
		\citenamefont {Ferrera}, \citenamefont {Faccio},\ and\ \citenamefont
		{Kinsey}}]{KH20}%
	\BibitemOpen
	\bibfield  {author} {\bibinfo {author} {\bibfnamefont {J.~B.}\ \bibnamefont
			{Khurgin}}, \bibinfo {author} {\bibfnamefont {M.}~\bibnamefont {Clerici}},
		\bibinfo {author} {\bibfnamefont {V.}~\bibnamefont {Bruno}}, \bibinfo
		{author} {\bibfnamefont {L.}~\bibnamefont {Caspani}}, \bibinfo {author}
		{\bibfnamefont {C.}~\bibnamefont {DeVault}}, \bibinfo {author} {\bibfnamefont
			{J.}~\bibnamefont {Kim}}, \bibinfo {author} {\bibfnamefont {A.}~\bibnamefont
			{Shaltout}}, \bibinfo {author} {\bibfnamefont {A.}~\bibnamefont
			{Boltasseva}}, \bibinfo {author} {\bibfnamefont {V.~M.}\ \bibnamefont
			{Shalaev}}, \bibinfo {author} {\bibfnamefont {M.}~\bibnamefont {Ferrera}},
		\bibinfo {author} {\bibfnamefont {D.}~\bibnamefont {Faccio}},\ and\ \bibinfo
		{author} {\bibfnamefont {N.}~\bibnamefont {Kinsey}},\ }\bibfield  {title}
	{\bibinfo {title} {Adiabatic frequency shifting in epsilon-near-zero
			materials: the role of group velocity},\ }\href@noop {} {\bibfield  {journal}
		{\bibinfo  {journal} {Optica}\ }\textbf {\bibinfo {volume} {7}},\ \bibinfo
		{pages} {226} (\bibinfo {year} {2020})}\BibitemShut {NoStop}%
	\bibitem [{\citenamefont {Zhou}\ \emph {et~al.}(2020)\citenamefont {Zhou},
		\citenamefont {Alam}, \citenamefont {Karimi}, \citenamefont {Upham},
		\citenamefont {Reshef}, \citenamefont {Liu}, \citenamefont {Willner},\ and\
		\citenamefont {Boyd}}]{ZH20}%
	\BibitemOpen
	\bibfield  {author} {\bibinfo {author} {\bibfnamefont {Y.}~\bibnamefont
			{Zhou}}, \bibinfo {author} {\bibfnamefont {Z.~A.}\ \bibnamefont {Alam}},
		\bibinfo {author} {\bibfnamefont {M.}~\bibnamefont {Karimi}}, \bibinfo
		{author} {\bibfnamefont {J.}~\bibnamefont {Upham}}, \bibinfo {author}
		{\bibfnamefont {O.}~\bibnamefont {Reshef}}, \bibinfo {author} {\bibfnamefont
			{C.}~\bibnamefont {Liu}}, \bibinfo {author} {\bibfnamefont {A.~E.}\
			\bibnamefont {Willner}},\ and\ \bibinfo {author} {\bibfnamefont {R.~W.}\
			\bibnamefont {Boyd}},\ }\bibfield  {title} {\bibinfo {title} {Broadband
			frequency translation through time refraction in an epsilon-near-zero
			material},\ }\href@noop {} {\bibfield  {journal} {\bibinfo  {journal} {Nat.
				Comm.}\ }\textbf {\bibinfo {volume} {11}},\ \bibinfo {pages} {2180} (\bibinfo
		{year} {2020})}\BibitemShut {NoStop}%
	\bibitem [{\citenamefont {Lustig}\ \emph {et~al.}(2023)\citenamefont {Lustig},
		\citenamefont {Segal}, \citenamefont {Saha}, \citenamefont {Bordo},
		\citenamefont {Chowdhury}, \citenamefont {Sharabi}, \citenamefont
		{Fleischer}, \citenamefont {Boltasseva}, \citenamefont {Cohen}, \citenamefont
		{Shalaev},\ and\ \citenamefont {Segev}}]{LU23}%
	\BibitemOpen
	\bibfield  {author} {\bibinfo {author} {\bibfnamefont {E.}~\bibnamefont
			{Lustig}}, \bibinfo {author} {\bibfnamefont {O.}~\bibnamefont {Segal}},
		\bibinfo {author} {\bibfnamefont {S.}~\bibnamefont {Saha}}, \bibinfo {author}
		{\bibfnamefont {E.}~\bibnamefont {Bordo}}, \bibinfo {author} {\bibfnamefont
			{S.~N.}\ \bibnamefont {Chowdhury}}, \bibinfo {author} {\bibfnamefont
			{Y.}~\bibnamefont {Sharabi}}, \bibinfo {author} {\bibfnamefont
			{A.}~\bibnamefont {Fleischer}}, \bibinfo {author} {\bibfnamefont
			{A.}~\bibnamefont {Boltasseva}}, \bibinfo {author} {\bibfnamefont
			{O.}~\bibnamefont {Cohen}}, \bibinfo {author} {\bibfnamefont {V.~M.}\
			\bibnamefont {Shalaev}},\ and\ \bibinfo {author} {\bibfnamefont
			{M.}~\bibnamefont {Segev}},\ }\bibfield  {title} {\bibinfo {title}
		{Time-refraction optics with single cycle modulation},\ }\href@noop {}
	{\bibfield  {journal} {\bibinfo  {journal} {Nanophotonics}\ }\textbf
		{\bibinfo {volume} {12}},\ \bibinfo {pages} {2221} (\bibinfo {year}
		{2023})}\BibitemShut {NoStop}%
	\bibitem [{\citenamefont {Naik}\ \emph {et~al.}(2011)\citenamefont {Naik},
		\citenamefont {Kim},\ and\ \citenamefont {Boltasseva}}]{NA11}%
	\BibitemOpen
	\bibfield  {author} {\bibinfo {author} {\bibfnamefont {G.~V.}\ \bibnamefont
			{Naik}}, \bibinfo {author} {\bibfnamefont {J.}~\bibnamefont {Kim}},\ and\
		\bibinfo {author} {\bibfnamefont {A.}~\bibnamefont {Boltasseva}},\ }\bibfield
	{title} {\bibinfo {title} {Oxides and nitrides as alternative plasmonic
			materials in the optical range [invited]},\ }\href
	{https://doi.org/10.1364/ome.1.001090} {\bibfield  {journal} {\bibinfo
			{journal} {Opt. Mater. Express}\ }\textbf {\bibinfo {volume} {1}},\ \bibinfo
		{pages} {1090} (\bibinfo {year} {2011})}\BibitemShut {NoStop}%
	\bibitem [{\citenamefont {Marqués}\ \emph {et~al.}(2002)\citenamefont
		{Marqués}, \citenamefont {Martel}, \citenamefont {Mesa},\ and\ \citenamefont
		{Medina}}]{MA02H}%
	\BibitemOpen
	\bibfield  {author} {\bibinfo {author} {\bibfnamefont {R.}~\bibnamefont
			{Marqués}}, \bibinfo {author} {\bibfnamefont {J.}~\bibnamefont {Martel}},
		\bibinfo {author} {\bibfnamefont {F.}~\bibnamefont {Mesa}},\ and\ \bibinfo
		{author} {\bibfnamefont {F.}~\bibnamefont {Medina}},\ }\bibfield  {title}
	{\bibinfo {title} {{Left-Handed-Media Simulation and Transmission of EM Waves
				in Subwavelength Split-Ring-Resonator-Loaded Metallic Waveguides}},\ }\href
	{https://doi.org/10.1103/physrevlett.89.183901} {\bibfield  {journal}
		{\bibinfo  {journal} {Phys. Rev. Lett.}\ }\textbf {\bibinfo {volume} {89}},\
		\bibinfo {pages} {183901} (\bibinfo {year} {2002})}\BibitemShut {NoStop}%
	\bibitem [{\citenamefont {Park}\ \emph {et~al.}(2015)\citenamefont {Park},
		\citenamefont {Kang}, \citenamefont {Liu},\ and\ \citenamefont
		{Brongersma}}]{PA15}%
	\BibitemOpen
	\bibfield  {author} {\bibinfo {author} {\bibfnamefont {J.}~\bibnamefont
			{Park}}, \bibinfo {author} {\bibfnamefont {J.-H.}\ \bibnamefont {Kang}},
		\bibinfo {author} {\bibfnamefont {X.}~\bibnamefont {Liu}},\ and\ \bibinfo
		{author} {\bibfnamefont {M.~L.}\ \bibnamefont {Brongersma}},\ }\bibfield
	{title} {\bibinfo {title} {{Electrically Tunable Epsilon-Near-Zero (ENZ)
				Metafilm Absorbers}},\ }\href@noop {} {\bibfield  {journal} {\bibinfo
			{journal} {Sci. Rep.}\ }\textbf {\bibinfo {volume} {5}},\ \bibinfo {pages}
		{15754} (\bibinfo {year} {2015})}\BibitemShut {NoStop}%
	\bibitem [{\citenamefont {Bohn}\ \emph {et~al.}(2021)\citenamefont {Bohn},
		\citenamefont {Luk}, \citenamefont {Tollerton}, \citenamefont {Hutchings},
		\citenamefont {Brener}, \citenamefont {Horsley}, \citenamefont {Barnes},\
		and\ \citenamefont {Hendry}}]{BO21}%
	\BibitemOpen
	\bibfield  {author} {\bibinfo {author} {\bibfnamefont {J.}~\bibnamefont
			{Bohn}}, \bibinfo {author} {\bibfnamefont {T.~S.}\ \bibnamefont {Luk}},
		\bibinfo {author} {\bibfnamefont {C.}~\bibnamefont {Tollerton}}, \bibinfo
		{author} {\bibfnamefont {S.~W.}\ \bibnamefont {Hutchings}}, \bibinfo {author}
		{\bibfnamefont {I.}~\bibnamefont {Brener}}, \bibinfo {author} {\bibfnamefont
			{S.}~\bibnamefont {Horsley}}, \bibinfo {author} {\bibfnamefont {W.~L.}\
			\bibnamefont {Barnes}},\ and\ \bibinfo {author} {\bibfnamefont
			{E.}~\bibnamefont {Hendry}},\ }\bibfield  {title} {\bibinfo {title}
		{All-optical switching of an epsilon-near-zero plasmon resonance in indium
			tin oxide},\ }\href@noop {} {\bibfield  {journal} {\bibinfo  {journal} {Nat.
				Comm.}\ }\textbf {\bibinfo {volume} {12}},\ \bibinfo {pages} {1017} (\bibinfo
		{year} {2021})}\BibitemShut {NoStop}%
	\bibitem [{\citenamefont {Garrett}\ and\ \citenamefont
		{McCumber}(1970)}]{GA70}%
	\BibitemOpen
	\bibfield  {author} {\bibinfo {author} {\bibfnamefont {C.~G.~B.}\
			\bibnamefont {Garrett}}\ and\ \bibinfo {author} {\bibfnamefont {D.~E.}\
			\bibnamefont {McCumber}},\ }\bibfield  {title} {\bibinfo {title} {Propagation
			of a gaussian light pulse through an anomalous dispersion medium},\ }\href
	{https://doi.org/10.1103/physreva.1.305} {\bibfield  {journal} {\bibinfo
			{journal} {Phys. Rev. A}\ }\textbf {\bibinfo {volume} {1}},\ \bibinfo {pages}
		{305} (\bibinfo {year} {1970})}\BibitemShut {NoStop}%
	\bibitem [{\citenamefont {Segard}\ and\ \citenamefont {Macke}(1985)}]{SE85}%
	\BibitemOpen
	\bibfield  {author} {\bibinfo {author} {\bibfnamefont {B.}~\bibnamefont
			{Segard}}\ and\ \bibinfo {author} {\bibfnamefont {B.}~\bibnamefont {Macke}},\
	}\bibfield  {title} {\bibinfo {title} {Observation of negative velocity pulse
			propagation},\ }\href {https://doi.org/10.1016/0375-9601(85)90305-6}
	{\bibfield  {journal} {\bibinfo  {journal} {Phys. Lett.}\ }\textbf {\bibinfo
			{volume} {109}},\ \bibinfo {pages} {213} (\bibinfo {year}
		{1985})}\BibitemShut {NoStop}%
	\bibitem [{\citenamefont {Chu}\ and\ \citenamefont {Wong}(1982)}]{CH82}%
	\BibitemOpen
	\bibfield  {author} {\bibinfo {author} {\bibfnamefont {S.}~\bibnamefont
			{Chu}}\ and\ \bibinfo {author} {\bibfnamefont {S.}~\bibnamefont {Wong}},\
	}\bibfield  {title} {\bibinfo {title} {Linear pulse propagation in an
			absorbing medium},\ }\href {https://doi.org/10.1103/physrevlett.48.738}
	{\bibfield  {journal} {\bibinfo  {journal} {Phys. Rev. Lett.}\ }\textbf
		{\bibinfo {volume} {48}},\ \bibinfo {pages} {738} (\bibinfo {year}
		{1982})}\BibitemShut {NoStop}%
		\bibitem [{\citenamefont {Turi}\ and\ \citenamefont {Rossky}(2012)}]{TU12B}%
	\BibitemOpen
	\bibfield  {author} {\bibinfo {author} {\bibfnamefont {L.}~\bibnamefont
			{Turi}}\ and\ \bibinfo {author} {\bibfnamefont {P.~J.}\ \bibnamefont
			{Rossky}},\ }\bibfield  {title} {\bibinfo {title} {{Theoretical Studies of
				Spectroscopy and Dynamics of Hydrated Electrons}},\ }\href
	{https://doi.org/10.1021/cr300144z} {\bibfield  {journal} {\bibinfo
			{journal} {Chem. Rev.}\ }\textbf {\bibinfo {volume} {112}},\ \bibinfo {pages}
		{5641} (\bibinfo {year} {2012})}\BibitemShut {NoStop}%
	\bibitem [{\citenamefont {Webster}(1972)}]{WE72}%
	\BibitemOpen
	\bibfield  {author} {\bibinfo {author} {\bibfnamefont {B.~C.}\ \bibnamefont
			{Webster}},\ }\bibfield  {title} {\bibinfo {title} {The polaron viewpoint of
			solvated electrons},\ }\href {https://doi.org/10.1038/physci239079a0}
	{\bibfield  {journal} {\bibinfo  {journal} {Nat. Phys. Sci.}\ }\textbf
		{\bibinfo {volume} {239}},\ \bibinfo {pages} {79} (\bibinfo {year}
		{1972})}\BibitemShut {NoStop}%
	\bibitem [{\citenamefont {Kenney-Wallace}\ and\ \citenamefont
		{Jonah}(1982)}]{KE82}%
	\BibitemOpen
	\bibfield  {author} {\bibinfo {author} {\bibfnamefont {G.~A.}\ \bibnamefont
			{Kenney-Wallace}}\ and\ \bibinfo {author} {\bibfnamefont {C.~D.}\
			\bibnamefont {Jonah}},\ }\bibfield  {title} {\bibinfo {title} {Picosecond
			spectroscopy and solvation clusters. {The} dynamics of localizing electrons
			in polar fluids},\ }\href {https://doi.org/10.1021/j100211a007} {\bibfield
		{journal} {\bibinfo  {journal} {J. Phys. Chem.}\ }\textbf {\bibinfo {volume}
			{86}},\ \bibinfo {pages} {2572} (\bibinfo {year} {1982})}\BibitemShut
	{NoStop}%
	\bibitem [{\citenamefont {Shi}\ \emph {et~al.}(1995)\citenamefont {Shi},
		\citenamefont {Long}, \citenamefont {Lu},\ and\ \citenamefont
		{Eisenthal}}]{SH95}%
	\BibitemOpen
	\bibfield  {author} {\bibinfo {author} {\bibfnamefont {X.}~\bibnamefont
			{Shi}}, \bibinfo {author} {\bibfnamefont {F.~H.}\ \bibnamefont {Long}},
		\bibinfo {author} {\bibfnamefont {H.}~\bibnamefont {Lu}},\ and\ \bibinfo
		{author} {\bibfnamefont {K.~B.}\ \bibnamefont {Eisenthal}},\ }\bibfield
	{title} {\bibinfo {title} {Electron solvation in neat alcohols},\ }\href
	{https://doi.org/10.1021/j100018a024} {\bibfield  {journal} {\bibinfo
			{journal} {J. Phys. Chem.}\ }\textbf {\bibinfo {volume} {99}},\ \bibinfo
		{pages} {6917} (\bibinfo {year} {1995})}\BibitemShut {NoStop}%
	\bibitem [{\citenamefont {Taub}\ \emph {et~al.}(1964)\citenamefont {Taub},
		\citenamefont {Harter}, \citenamefont {Sauer},\ and\ \citenamefont
		{Dorfman}}]{TA64}%
	\BibitemOpen
	\bibfield  {author} {\bibinfo {author} {\bibfnamefont {I.~A.}\ \bibnamefont
			{Taub}}, \bibinfo {author} {\bibfnamefont {D.~A.}\ \bibnamefont {Harter}},
		\bibinfo {author} {\bibfnamefont {M.~C.}\ \bibnamefont {Sauer}},\ and\
		\bibinfo {author} {\bibfnamefont {L.~M.}\ \bibnamefont {Dorfman}},\
	}\bibfield  {title} {\bibinfo {title} {Pulse radiolysis studies. {IV}. the
			solvated electron in the aliphatic alcohols},\ }\href
	{https://doi.org/10.1063/1.1726042} {\bibfield  {journal} {\bibinfo
			{journal} {J. Chem. Phys.}\ }\textbf {\bibinfo {volume} {41}},\ \bibinfo
		{pages} {979} (\bibinfo {year} {1964})}\BibitemShut {NoStop}%
	\bibitem [{\citenamefont {Savolainen}\ \emph {et~al.}(2014)\citenamefont
		{Savolainen}, \citenamefont {Uhlig}, \citenamefont {Ahmed}, \citenamefont
		{Hamm},\ and\ \citenamefont {Jungwirth}}]{SA14}%
	\BibitemOpen
	\bibfield  {author} {\bibinfo {author} {\bibfnamefont {J.}~\bibnamefont
			{Savolainen}}, \bibinfo {author} {\bibfnamefont {F.}~\bibnamefont {Uhlig}},
		\bibinfo {author} {\bibfnamefont {S.}~\bibnamefont {Ahmed}}, \bibinfo
		{author} {\bibfnamefont {P.}~\bibnamefont {Hamm}},\ and\ \bibinfo {author}
		{\bibfnamefont {P.}~\bibnamefont {Jungwirth}},\ }\bibfield  {title} {\bibinfo
		{title} {Direct observation of the collapse of the delocalized excess
			electron in water},\ }\href {https://doi.org/10.1038/nchem.1995} {\bibfield
		{journal} {\bibinfo  {journal} {Nat. Chem.}\ }\textbf {\bibinfo {volume}
			{6}},\ \bibinfo {pages} {697} (\bibinfo {year} {2014})}\BibitemShut {NoStop}%
	\bibitem [{\citenamefont {Ghalgaoui}\ \emph {et~al.}(2020)\citenamefont
		{Ghalgaoui}, \citenamefont {Koll}, \citenamefont {Sch{\"u}tte}, \citenamefont
		{Fingerhut}, \citenamefont {Reimann}, \citenamefont {Woerner},\ and\
		\citenamefont {Elsaesser}}]{GH20A}%
	\BibitemOpen
	\bibfield  {author} {\bibinfo {author} {\bibfnamefont {A.}~\bibnamefont
			{Ghalgaoui}}, \bibinfo {author} {\bibfnamefont {L.-M.}\ \bibnamefont {Koll}},
		\bibinfo {author} {\bibfnamefont {B.}~\bibnamefont {Sch{\"u}tte}}, \bibinfo
		{author} {\bibfnamefont {B.~P.}\ \bibnamefont {Fingerhut}}, \bibinfo {author}
		{\bibfnamefont {K.}~\bibnamefont {Reimann}}, \bibinfo {author} {\bibfnamefont
			{M.}~\bibnamefont {Woerner}},\ and\ \bibinfo {author} {\bibfnamefont
			{T.}~\bibnamefont {Elsaesser}},\ }\bibfield  {title} {\bibinfo {title}
		{Field-induced tunneling ionization and terahertz-driven electron dynamics in
			liquid water},\ }\href {https://doi.org/10.1021/acs.jpclett.0c02312}
	{\bibfield  {journal} {\bibinfo  {journal} {J. Phys. Chem. Lett.}\ }\textbf
		{\bibinfo {volume} {11}},\ \bibinfo {pages} {7717} (\bibinfo {year}
		{2020})}\BibitemShut {NoStop}%
	\bibitem [{\citenamefont {Ghalgaoui}\ \emph {et~al.}(2021)\citenamefont
		{Ghalgaoui}, \citenamefont {Fingerhut}, \citenamefont {Reimann},
		\citenamefont {Elsaesser},\ and\ \citenamefont {Woerner}}]{GH21B}%
	\BibitemOpen
	\bibfield  {author} {\bibinfo {author} {\bibfnamefont {A.}~\bibnamefont
			{Ghalgaoui}}, \bibinfo {author} {\bibfnamefont {B.~P.}\ \bibnamefont
			{Fingerhut}}, \bibinfo {author} {\bibfnamefont {K.}~\bibnamefont {Reimann}},
		\bibinfo {author} {\bibfnamefont {T.}~\bibnamefont {Elsaesser}},\ and\
		\bibinfo {author} {\bibfnamefont {M.}~\bibnamefont {Woerner}},\ }\bibfield
	{title} {\bibinfo {title} {Terahertz polaron oscillations of electrons
			solvated in liquid water},\ }\href
	{https://doi.org/10.1103/PhysRevLett.126.097401} {\bibfield  {journal}
		{\bibinfo  {journal} {Phys. Rev. Lett.}\ }\textbf {\bibinfo {volume} {126}},\
		\bibinfo {pages} {097401} (\bibinfo {year} {2021})}\BibitemShut {NoStop}%
	\bibitem [{\citenamefont {Singh}\ \emph {et~al.}(2022)\citenamefont {Singh},
		\citenamefont {Zhang}, \citenamefont {Ghalgaoui}, \citenamefont {Reimann},
		\citenamefont {Fingerhut}, \citenamefont {Woerner},\ and\ \citenamefont
		{Elsaesser}}]{SI22B}%
	\BibitemOpen
	\bibfield  {author} {\bibinfo {author} {\bibfnamefont {P.}~\bibnamefont
			{Singh}}, \bibinfo {author} {\bibfnamefont {J.}~\bibnamefont {Zhang}},
		\bibinfo {author} {\bibfnamefont {A.}~\bibnamefont {Ghalgaoui}}, \bibinfo
		{author} {\bibfnamefont {K.}~\bibnamefont {Reimann}}, \bibinfo {author}
		{\bibfnamefont {B.~P.}\ \bibnamefont {Fingerhut}}, \bibinfo {author}
		{\bibfnamefont {M.}~\bibnamefont {Woerner}},\ and\ \bibinfo {author}
		{\bibfnamefont {T.}~\bibnamefont {Elsaesser}},\ }\bibfield  {title} {\bibinfo
		{title} {{Coherent polaron dynamics of electrons solvated in polar
				liquids}},\ }\href {https://doi.org/10.1093/pnasnexus/pgac078} {\bibfield
		{journal} {\bibinfo  {journal} {PNAS Nexus}\ }\textbf {\bibinfo {volume}
			{1}},\ \bibinfo {pages} {pgac078} (\bibinfo {year} {2022})}\BibitemShut
	{NoStop}%
	\bibitem [{\citenamefont {Runge}\ \emph {et~al.}(2023)\citenamefont {Runge},
		\citenamefont {Reimann}, \citenamefont {Woerner},\ and\ \citenamefont
		{Elsaesser}}]{RU23A}%
	\BibitemOpen
	\bibfield  {author} {\bibinfo {author} {\bibfnamefont {M.}~\bibnamefont
			{Runge}}, \bibinfo {author} {\bibfnamefont {K.}~\bibnamefont {Reimann}},
		\bibinfo {author} {\bibfnamefont {M.}~\bibnamefont {Woerner}},\ and\ \bibinfo
		{author} {\bibfnamefont {T.}~\bibnamefont {Elsaesser}},\ }\bibfield  {title}
	{\bibinfo {title} {Nonlinear terahertz polarizability of electrons solvated
			in a polar liquid},\ }\href {https://doi.org/10.1103/physrevlett.131.166902}
	{\bibfield  {journal} {\bibinfo  {journal} {Phys. Rev. Lett.}\ }\textbf
		{\bibinfo {volume} {131}},\ \bibinfo {pages} {166902} (\bibinfo {year}
		{2023})}\BibitemShut {NoStop}%
\bibitem [{\citenamefont {Hannay}\ (2022)\citenamefont {Hannay},}]{HA83}%
	\BibitemOpen
	\bibfield  {author} {\bibinfo {author} {\bibfnamefont {J. H.}~\bibnamefont
			{Hannay}},\ }\bibfield  {title}
	{\bibinfo {title} {The Clausius-Mossotti equation: An alternative derivation},\ }
	\href {https://doi.org/10.1088/0143-0807/4/3/003}
	{\bibfield  {journal} {\bibinfo  {journal} {Eur. J. Phys.}\ }\textbf
		{\bibinfo {volume} {4}},\ \bibinfo {pages} {141-143} (\bibinfo {year}
		{1883})}\BibitemShut {NoStop}%
\bibitem [{\citenamefont {Woerner}\ \emph {et~al.}(2022)\citenamefont {Woerner},
		\citenamefont {Fingerhut}, \ and\ \citenamefont {Elsaesser}}]{WO22}%
	\BibitemOpen
	\bibfield  {author} {\bibinfo {author} {\bibfnamefont {M}~\bibnamefont
			{Woerner}}, \bibinfo {author} {\bibfnamefont {B. P.}~\bibnamefont {Fingerhut}},
		\ and\ \bibinfo
		{author} {\bibfnamefont {T.}~\bibnamefont {Elsaesser}},\ }\bibfield  {title}
	{\bibinfo {title} {Field-induced electron generation in water: Solvation dynamics and many-body interactions},\ }
	\href {https://doi.org/10.1021/acs.jpcb.2c01102}
	{\bibfield  {journal} {\bibinfo  {journal} {J. Phys. Chem. B}\ }\textbf
		{\bibinfo {volume} {126}},\ \bibinfo {pages} {2621-2634} (\bibinfo {year}
		{2022})}\BibitemShut {NoStop}%
		\bibitem [{\citenamefont {Sauer, Jr.}\ \emph {et~al.}(2023)\citenamefont {Sauer, Jr.},
		\citenamefont {Arai}, \ and\ \citenamefont {Dorfman}}]{SA65}%
	\BibitemOpen
	\bibfield  {author} {\bibinfo {author} {\bibfnamefont {M. C.}~\bibnamefont
			{Sauer, Jr.}}, \bibinfo {author} {\bibfnamefont {S.}~\bibnamefont {Arai}},
		\ and\ \bibinfo
		{author} {\bibfnamefont {L. M.}~\bibnamefont {Webster}},\ }\bibfield  {title}
	{\bibinfo {title} {Pulse radiolysis studies. VII. The absorption spectra and radiation chemical yields of the solvated electron
			in the aliphatic alcohols},\ }\href {https://doi.org/10.1063/1.1695994}
	{\bibfield  {journal} {\bibinfo  {journal} {J. Chem. Phys.}\ }\textbf
		{\bibinfo {volume} {42}},\ \bibinfo {pages} {708--712} (\bibinfo {year}
		{1965})}\BibitemShut {NoStop}%
	\bibitem [{\citenamefont {Reimann}\ \emph {et~al.}(2021)\citenamefont
		{Reimann}, \citenamefont {Woerner},\ and\ \citenamefont {Elsaesser}}]{RE21}%
	\BibitemOpen
	\bibfield  {author} {\bibinfo {author} {\bibfnamefont {K.}~\bibnamefont
			{Reimann}}, \bibinfo {author} {\bibfnamefont {M.}~\bibnamefont {Woerner}},\
		and\ \bibinfo {author} {\bibfnamefont {T.}~\bibnamefont {Elsaesser}},\
	}\bibfield  {title} {\bibinfo {title} {Two-dimensional terahertz spectroscopy
			of condensed-phase molecular systems},\ }\href@noop {} {\bibfield  {journal}
		{\bibinfo  {journal} {J. Chem. Phys.}\ }\textbf {\bibinfo {volume} {154}},\
		\bibinfo {pages} {120901} (\bibinfo {year} {2021})}\BibitemShut {NoStop}%
	\bibitem [{\citenamefont {Shih}\ \emph {et~al.}(2005)\citenamefont {Shih},
		\citenamefont {Reimann}, \citenamefont {Woerner}, \citenamefont {Elsaesser},
		\citenamefont {Waldm{\"u}ller}, \citenamefont {Knorr}, \citenamefont {Hey},\
		and\ \citenamefont {Ploog}}]{SH05A}%
	\BibitemOpen
	\bibfield  {author} {\bibinfo {author} {\bibfnamefont {T.}~\bibnamefont
			{Shih}}, \bibinfo {author} {\bibfnamefont {K.}~\bibnamefont {Reimann}},
		\bibinfo {author} {\bibfnamefont {M.}~\bibnamefont {Woerner}}, \bibinfo
		{author} {\bibfnamefont {T.}~\bibnamefont {Elsaesser}}, \bibinfo {author}
		{\bibfnamefont {I.}~\bibnamefont {Waldm{\"u}ller}}, \bibinfo {author}
		{\bibfnamefont {A.}~\bibnamefont {Knorr}}, \bibinfo {author} {\bibfnamefont
			{R.}~\bibnamefont {Hey}},\ and\ \bibinfo {author} {\bibfnamefont {K.~H.}\
			\bibnamefont {Ploog}},\ }\bibfield  {title} {\bibinfo {title} {{Nonlinear
				response of radiatively coupled intersubband transitions of
				quasi--two-dimensional electrons}},\ }\href
	{https://doi.org/10.1103/PhysRevB.72.195338} {\bibfield  {journal} {\bibinfo
			{journal} {Phys. Rev. B}\ }\textbf {\bibinfo {volume} {72}},\ \bibinfo
		{pages} {195338} (\bibinfo {year} {2005})}\BibitemShut {NoStop}%
	\bibitem [{\citenamefont {Masur}\ \emph {et~al.}()\citenamefont {Masur},
		\citenamefont {Bondar},\ and\ \citenamefont {McCaul}}]{MA23}%
	\BibitemOpen
	\bibfield  {author} {\bibinfo {author} {\bibfnamefont {J.}~\bibnamefont
			{Masur}}, \bibinfo {author} {\bibfnamefont {D.~I.}\ \bibnamefont {Bondar}},\
		and\ \bibinfo {author} {\bibfnamefont {G.}~\bibnamefont {McCaul}},\
	}\bibfield  {title} {\bibinfo {title} {Dynamical generation of
			epsilon-near-zero behaviour via tracking and feedback control},\ }\href@noop
	{} {\bibinfo  {journal} {arXiv:2301.12069}\ }\BibitemShut {NoStop}%
\end{thebibliography}

\ifx\unpublished\@undefined\def\unpublished{unpublished}\fi

\end{document}